\documentclass[pra,12pt,tightenlines]{revtex4}
\usepackage{latexsym}
\usepackage{graphicx}
\usepackage{amsmath}
\usepackage{amsbsy}
\usepackage{amsthm}
\usepackage{bbm}
\usepackage{bm}
\newtheorem{thm}{Theorem}
\newtheorem{lem}{Lemma}
\newtheorem{defi}{Definition}

\begin{document}

\title{Bloch vectors for qudits}
\author{Reinhold A. Bertlmann} \email{reinhold.bertlmann@univie.ac.at}
\author{Philipp Krammer} \email{philipp.krammer@univie.ac.at}
\affiliation{Faculty of Physics, University of Vienna, Boltzmanngasse 5, A-1090 Vienna,
Austria}

\begin{abstract}

We present three different matrix bases that can be used to decompose density matrices of
$d$--dimensional quantum systems, so-called qudits: the \emph{generalized Gell-Mann
matrix basis}, the \emph{polarization operator basis}, and the \emph{Weyl operator
basis}. Such a decomposition can be identified with a vector ---the Bloch vector, i.e. a
generalization of the well known qubit case--- and is a convenient expression for
comparison with measurable quantities and for explicit calculations avoiding the handling
of large matrices. We present a new method to decompose density matrices via so--called
standard matrices, consider the important case of an isotropic two--qudit state and
decompose it according to each basis. In case of qutrits we show a representation of
an entanglement witness in terms of expectation values of spin 1 measurements, which
is appropriate for an experimental realization.

\end{abstract}

\maketitle

\section{Introduction}

The state of a $d$--dimensional quantum system ---a qudit--- is usually described by a $d
\times d$ density matrix. For high dimensions, where the matrices become large (for
composite systems of $n$ particles the matrices are of even much larger dimension
$d^{\,n} \times d^{\,n}$), a simple way to express density matrices is of great interest.

Since the space of matrices is a vector space, there exist bases of matrices which can be
used to decompose any matrix. For qubits such a basis contains the three Pauli matrices,
accordingly, a density matrix can be expressed by a 3--dimensional vector, the
\emph{Bloch vector}, and any such vector has to lie within the so-called \emph{Bloch
ball} \cite{bloch46, nielsen00}. Unique for qubits is the fact that any point on the
sphere, \emph{Bloch sphere}, and inside the ball corresponds to a physical state, i.e. a
density matrix. The pure states lie on the sphere and the mixed ones inside.

In higher dimensions there exist different matrix bases that can be used to express
qudits as ($d^{\,2}-1$)--dimensional vectors as well. Different to the qubit case,
however, is that the map induced is not bijective: not every point on the ``Bloch
sphere'' in dimensions $d^{\,2}-1$ corresponds to a physical state. Nevertheless the
vectors are often also called ``Bloch vectors'' (see in this context, e.g.,
Refs.~\cite{kimura03, kryszewski06, jakobczyk01, kimura05, mendas06}).

In this paper we want to present and compare three different matrix bases for a Bloch
vector decomposition of qudits. In Sec.~\ref{prelim} we propose the properties of any
matrix basis for using it as a ``practical'' decomposition of density matrices and recall
the general notation of Bloch vectors. In Secs.~\ref{secggb} -- \ref{secwob} we offer
three different matrix bases: the \emph{generalized Gell-Mann matrix basis}, the
\emph{polarization operator basis}, and the \emph{Weyl operator basis}. For all these
bases we give examples in the dimensions of our interest and present the different Bloch
vector decompositions of an arbitrary density matrix in the standard matrix notation.
Next in Sec.~\ref{seciso}, by constructing tensor products of states we study the
isotropic two--qudit state and present the results for the three matrix decompositions,
i.e. for the three different Bloch vectors. In Sec.~\ref{secapp} we focus on the
isotropic two--qudit state and calculate the Hilbert--Schmidt measure of entanglement
(see, e.g., Refs.~\cite{witte99, ozawa00, bertlmann02, bertlmann05}). Its connection to
the optimal entanglement witness is shown, which is determined in terms of the three
matrix bases. An example for the experimental realization of an entanglement witness is
given in Sec.~\ref{witness-experimental}. The mathematical and physical
advantages/disadvantages by using the three different matrix bases are discussed in
Sec.~\ref{conclusion}, where also the final conclusions are drawn.

\section{Preliminaries}
\label{prelim}

A \emph{qudit} state is represented by a density operator in the Hilbert--Schmidt space acting on
the d--dimensional Hilbert space ${\cal H}^{\,d}$ that can be written as a matrix ---the density
matrix--- in the \emph{standard basis} $ \left\{ \left| k
\right\rangle \right\}, \,$ with $k=1,2, \ldots d \,$ or $k=0,1,2, \ldots d-1$.\\

\emph{Properties of a ``practical'' matrix basis.} For practical reasons the general properties of
a matrix basis which is used for the Bloch vector decomposition of qudits are the following:
\begin{enumerate}
    \item[i)] The basis includes the identity matrix $\mathbbm{1}$ and $d-1$ matrices
    $\left\{ A_i \right\}$ of dimension $d \times d$ which are traceless, i.e. $\textnormal{Tr}
    A_i = 0 \,$.
    \item[ii)] The matrices of any basis $\left\{ A_i \right\}$ are orthogonal, i.e.
    \begin{equation} \label{orthogon}
    \textnormal{Tr}\, A_i^{\dag} A_j \,=\, N \, \delta_{ij} \quad \textrm{with} \quad N\in
    \mathbbm{R}\,.\\
    \end{equation}
\end{enumerate}

\emph{Bloch vector expansion of a density matrix.} Since any matrix in the
Hilbert-Schmidt space of dimension $d$ can be decomposed with a matrix basis $\left\{ A_i
\right\}$, we can of course decompose a qudit density matrix as well and get the
\emph{Bloch vector expansion} of the density matrix,
\begin{equation} \label{bvgenbasis}
    \rho \;=\; \frac{1}{d} \,\mathbbm{1} \,+\, \vec{b} \cdot \vec{\Gamma} \,,
\end{equation}
where $\vec{b} \cdot\vec{\Gamma}$ is a linear combination of all matrices $\left\{ A_i \right\}$
and the vector $\vec{b} \in \mathbbm{R}^{d^2-1}$ with $b_i = \langle \Gamma_i \rangle =
\textnormal{Tr}\rho\Gamma_i\,$ is called \emph{Bloch vector}. The term
$\frac{1}{d} \mathbbm{1}$ is fixed because of condition $\text{Tr} \rho = 1$.\\

\emph{Remark.} Note that a given density matrix $\rho$ can always be decomposed into a
Bloch vector, but not any vector $\sigma$ that is of the form \eqref{bvgenbasis} is
automatically a density matrix, even if it satisfies the conditions Tr$\sigma = 1$ and
Tr$\sigma^2 \leq 1$ since generally it does not imply $\sigma \geq 0$.

Each different matrix basis induces a different Bloch vector lying within a Bloch
hypersphere where, however, not every point of the hypersphere corresponds to a physical
state (with $\rho \geq 0$); these points are excluded (holes). The geometric character of
the Bloch space in higher dimensions turns out to be quite complicated and is still of
great interest (see Refs.~\cite{kimura03, kryszewski06, jakobczyk01, kimura05,
mendas06}).

All different Bloch hyperballs are isomorphic since they correspond to the same density
matrix $\rho$. The interesting question is which Bloch hyperball ---which matrix basis---
is optimal for a specific purpose, like the calculation of the entanglement degree or the
determination of the geometry of the Hilbert space or the comparison with measurable
quantities.

\section{The Generalized Gell-Mann Matrix Basis}
\label{secggb}

\subsection{Definition and example}

The generalized Gell-Mann matrices (GGM) are higher--dimensional extensions of the Pauli
matrices (for qubits) and the Gell-Mann matrices (for qutrits), they are the standard
SU(N) generators (in our case $N=d$). They are defined as three different types of
matrices and for simplicity we use here the operator notation; then the density matrices
follow by simply writing the operators in the standard basis (see, e.g.
Refs.~\cite{kimura03, rakotonirina05}):
\begin{enumerate}
\item[i)]{$\frac{d(d-1)}{2}$ symmetric GGM}
    \begin{equation} \label{ggms}
        \Lambda^{jk}_s \;=\; | j \rangle \langle k | \,+\, | k \rangle
        \langle j |\,, \quad 1 \leq j < k \leq d \;,
    \end{equation}
\item[ii)]{$\frac{d(d-1)}{2}$ antisymmetric GGM}
    \begin{equation} \label{ggma}
        \Lambda^{jk}_a \;=\; -i \,| j \rangle \langle k | \,+\, i \,| k \rangle
        \langle j |\,, \quad 1 \leq j < k \leq d \;,
    \end{equation}
\item[iii)]{$(d-1)$ diagonal GGM}
    \begin{equation} \label{ggmd}
        \Lambda^l \;=\; \sqrt{\frac{2}{l(l+1)}} \left( \sum_{j=1}^{l} |
        j \rangle \langle j | \,-\, l \,| l+1 \rangle \langle l+1 |
        \right), \quad 1 \leq l \leq d-1 \;.
    \end{equation}
\end{enumerate}
In total we have $d^2-1$ GGM; it follows from the definitions that all GGM are Hermitian
and traceless. They are orthogonal and form a basis, the generalized Gell-Mann matrix
basis (GGB). A proof for the orthogonality of GGB we present in the
Appendix \ref{orthogonality}.\\

\emph{Examples.} Let us recall the case of dimension $3$, the $8$ Gell-Mann matrices (for
a representation see, e.g., Refs.~\cite{bertlmann05,caves00})
\begin{enumerate}
\item[i)]{$3$ symmetric Gell-Mann matrices}
\begin{eqnarray} \label{defgellmann-s}
    & \lambda^{12}_s = \left(
    \begin{array}{ccc}
        0 & 1 & 0 \\
        1 & 0 & 0 \\
        0 & 0 & 0 \\
    \end{array} \right), \quad
    \lambda^{13}_s = \left(
    \begin{array}{ccc}
        0 & 0 & 1 \\
        0 & 0 & 0 \\
        1 & 0 & 0 \\
    \end{array} \right), \quad
    \lambda^{23}_s = \left(
    \begin{array}{ccc}
        0 & 0 & 0 \\
        0 & 0 & 1 \\
        0 & 1 & 0 \\
    \end{array} \right), &
    \nonumber \\
\end{eqnarray}
\item[ii)]{$3$ antisymmetric Gell-Mann matrices}
\begin{eqnarray} \label{defgellmann-a}
    & \lambda^{12}_a = \left(
    \begin{array}{ccc}
        0 & -i & 0 \\
        i & 0 & 0 \\
        0 & 0 & 0 \\
    \end{array} \right), \quad
    \lambda^{13}_a = \left(
    \begin{array}{ccc}
        0 & 0 & -i \\
        0 & 0 & 0 \\
        i & 0 & 0 \\
    \end{array} \right), \quad
    \lambda^{23}_a = \left(
    \begin{array}{ccc}
        0 & 0 & 0 \\
        0 & 0 & -i \\
        0 & i & 0 \\
    \end{array} \right), &
    \nonumber \\
\end{eqnarray}
\item[ii)]{$2$ diagonal Gell-Mann matrices}
\begin{eqnarray} \label{defgellmann-d}
    & \lambda^1 = \left(
    \begin{array}{ccc}
        1 & 0 & 0 \\
        0 & -1 & 0 \\
        0 & 0 & 0 \\
    \end{array} \right), \quad
    \lambda^2 = \frac{1}{\sqrt{3}} \left(
    \begin{array}{ccc}
        1 & 0 & 0 \\
        0 & 1 & 0 \\
        0 & 0 & -2 \\
    \end{array} \right). &
\end{eqnarray}
\end{enumerate}

To see how they generalize for higher dimensions we show the case we need for qudits of
dimension $d=4\,$:
\begin{enumerate}
\item[i)]{$6$ symmetric GGM}
\begin{eqnarray} \label{ggm4s}
    & \Lambda^{12}_s = \left(
    \begin{array}{cccc}
        0 & 1 & 0 & 0 \\
        1 & 0 & 0 & 0 \\
        0 & 0 & 0 & 0 \\
        0 & 0 & 0 & 0 \\
    \end{array} \right), \quad
    \Lambda^{13}_s = \left(
    \begin{array}{cccc}
        0 & 0 & 1 & 0 \\
        0 & 0 & 0 & 0 \\
        1 & 0 & 0 & 0 \\
        0 & 0 & 0 & 0 \\
    \end{array} \right), \quad
    \Lambda^{14}_s = \left(
    \begin{array}{cccc}
        0 & 0 & 0 & 1 \\
        0 & 0 & 0 & 0 \\
        0 & 0 & 0 & 0 \\
        1 & 0 & 0 & 0 \\
    \end{array} \right), &
    \nonumber \\
    & \Lambda^{23}_s = \left(
    \begin{array}{cccc}
        0 & 0 & 0 & 0 \\
        0 & 0 & 1 & 0\\
        0 & 1 & 0 & 0 \\
        0 & 0 & 0 & 0 \\
    \end{array} \right), \quad
    \Lambda^{24}_s = \left(
    \begin{array}{cccc}
        0 & 0 & 0 & 0 \\
        0 & 0 & 0 & 1 \\
        0 & 0 & 0 & 0 \\
        0 & 1 & 0 & 0 \\
    \end{array} \right), \quad
    \Lambda^{34}_s = \left(
    \begin{array}{cccc}
        0 & 0 & 0 & 0 \\
        0 & 0 & 0 & 0 \\
        0 & 0 & 0 & 1 \\
        0 & 0 & 1 & 0 \\
    \end{array} \right), &
\end{eqnarray}
\item[ii)]{$6$ antisymmetric GGM}
\begin{eqnarray} \label{ggm4a}
    & \Lambda^{12}_a = \left(
    \begin{array}{cccc}
        0 & -i & 0 & 0 \\
        i & 0 & 0 & 0 \\
        0 & 0 & 0 & 0 \\
        0 & 0 & 0 & 0 \\
    \end{array} \right), \quad
    \Lambda^{13}_a = \left(
    \begin{array}{cccc}
        0 & 0 & -i & 0 \\
        0 & 0 & 0 & 0 \\
        i & 0 & 0 & 0 \\
        0 & 0 & 0 & 0 \\
    \end{array} \right), \quad
    \Lambda^{14}_a = \left(
    \begin{array}{cccc}
        0 & 0 & 0 & -i \\
        0 & 0 & 0 & 0 \\
        0 & 0 & 0 & 0 \\
        i & 0 & 0 & 0 \\
    \end{array} \right), &
    \nonumber \\
    & \Lambda^{23}_a = \left(
    \begin{array}{cccc}
        0 & 0 & 0 & 0 \\
        0 & 0 & -i & 0\\
        0 & i & 0 & 0 \\
        0 & 0 & 0 & 0 \\
    \end{array} \right), \quad
    \Lambda^{24}_a = \left(
    \begin{array}{cccc}
        0 & 0 & 0 & 0 \\
        0 & 0 & 0 & -i \\
        0 & 0 & 0 & 0 \\
        0 & i & 0 & 0 \\
    \end{array} \right), \quad
    \Lambda^{34}_a = \left(
    \begin{array}{cccc}
        0 & 0 & 0 & 0 \\
        0 & 0 & 0 & 0 \\
        0 & 0 & 0 & -i \\
        0 & 0 & i & 0 \\
    \end{array} \right), &
\end{eqnarray}
\item[iii)]{$3$ diagonal GGM}
\begin{equation} \label{ggm4d}
    \Lambda^{1} = \left(
    \begin{array}{cccc}
        1 & 0 & 0 & 0 \\
        0 & -1 & 0 & 0 \\
        0 & 0 & 0 & 0 \\
        0 & 0 & 0 & 0 \\
    \end{array} \right), \quad
    \Lambda^{2} = \frac{1}{\sqrt{3}} \left(
    \begin{array}{cccc}
        1 & 0 & 0 & 0 \\
        0 & 1 & 0 & 0 \\
        0 & 0 & -2 & 0 \\
        0 & 0 & 0 & 0 \\
    \end{array} \right), \quad
    \Lambda^{3} = \frac{1}{\sqrt{6}} \left(
    \begin{array}{cccc}
        1 & 0 & 0 & 0 \\
        0 & 1 & 0 & 0 \\
        0 & 0 & 1 & 0 \\
        0 & 0 & 0 & -3 \\
    \end{array} \right).
\end{equation}
\end{enumerate}

Using the GGB we obtain, in general, the following Bloch vector expansion of a density
matrix:
\begin{equation} \label{bvggb-d}
    \rho \;=\; \frac{1}{d} \,\mathbbm{1} \,+\, \vec{b} \cdot \vec{\Lambda} \,,
\end{equation}
with the Bloch vector $\vec{b} = \big(\{b^{jk}_s\},\{b^{jk}_a\},\{b^l\}\big)\,$, where
the components are ordered and for the indices we have the restrictions $1 \leq j < k
\leq d$ and $1 \leq l \leq d-1\,$. The components are given by $b^{jk}_s =
\textrm{Tr}\Lambda^{jk}_s\rho\,$, $b^{jk}_a = \textrm{Tr}\Lambda^{jk}_a\rho$ and $b^l =
\textrm{Tr}\Lambda^l\rho\,$. All Bloch vectors lie within a hypersphere of radius
$|\vec{b}| \leq \sqrt{(d-1)/2d}\,$. For example, for qutrits the Bloch vector components
are $\vec{b} = \big(b^{12}_s,b^{13}_s,b^{23}_s,b^{12}_a,b^{13}_a,b^{23}_a,b^1,b^2\big)$
corresponding to the Gell-Mann matrices \eqref{defgellmann-s}, \eqref{defgellmann-a},
\eqref{defgellmann-d} and $|\vec{b}| \leq \sqrt{1/3}\,$.

As already mentioned the allowed range of $\vec{b}$ is restricted. It has an interesting
geometric structure which has been calculated analytically for the case of qutrits by
studying $2$--dimensional planes in the $8$--dimensional Bloch space \cite{kimura03} or
numerically by considering $3$--dimensional cross--sections \cite{mendas06}. In any case,
pure states lie on the surface and the mixed ones inside.

\subsection{Standard matrix basis expansion by GGB}

The standard matrices are simply the $d \times d$ matrices that have only one entry $1$
and the other entries $0$ and form an orthonormal basis of the Hilbert--Schmidt space. We
write these matrices shortly as operators
\begin{equation} \label{standardmatrices}
    | j \rangle \langle k | \,, \qquad \textrm{with} \quad j,k = 1, \ldots, d \,.
\end{equation}
Any matrix can easily be decomposed into a ``vector'' via a certain linear combination of
the matrices \eqref{standardmatrices}. Knowing the expansion of matrices
\eqref{standardmatrices} into GGB we can therefore find the decomposition of any matrix
in terms of the GGB.

We find the following expansion of standard matrices
\eqref{standardmatrices} into GGB \,:
\begin{equation} \label{smggb}
    | j \rangle \langle k | \;=\; \begin{cases}
        \frac{1}{2} \left( \Lambda^{jk}_s \,+\, i \Lambda^{jk}_a \right)
            & \text{for } j < k \\
        \frac{1}{2} \left( \Lambda^{kj}_s \,-\, i \Lambda^{kj}_a \right)
            & \text{for } j > k \\
        - \sqrt{\frac{j-1}{2j}} \,\Lambda^{j-1} \,+\, \sum\limits_{n=0}^{d-j-1}
            \frac{1}{\sqrt{2(j+n)(j+n+1)}} \,\Lambda^{j+n} \,+\, \frac{1}{d}
            \,\mathbbm{1} & \text{for } j = k \,.
        \end{cases}
\end{equation}

\emph{Proof.} The first two cases can be easily verified.

To show the last case we first set up a recurrence relation for $| l \rangle \langle l|$, which we
obtain by eliminating the term $\sum_{j=1}^{l-1} | j \rangle \langle j|$ in the two expressions
\eqref{ggmd} for $\Lambda^l$ and $\Lambda^{l-1}\,$
\begin{equation} \label{recl}
    | l \rangle \langle l| \;=\; - \sqrt{\frac{l-1}{2l}} \,\Lambda^{l-1} \,+\,
    \sqrt{\frac{l+1}{2l}} \,\Lambda^l \,+\, |l+1\rangle \langle l+1| \,,
\end{equation}
and we consider the case $l+1=d$
\begin{equation} \label{recd}
| d-1 \rangle \langle d-1| \;=\; - \sqrt{\frac{d-2}{2(d-1)}} \,\Lambda^{d-2} \,+\,
    \sqrt{\frac{d}{2(d-1)}} \,\Lambda^{d-1} \,+\, |d\rangle \langle d| \,.
\end{equation}
From $\Lambda^{d-1}$ given by Eq.~\eqref{ggmd}
\begin{equation}
    \Lambda^{d-1} \;=\; \sqrt{\frac{2}{(d-1)d}} \,\left( \sum_{j=1}^{d-1}
    |j \rangle \langle j | \,-\, (d-1) |d \rangle \langle d| \right) \,,
\end{equation}
we get the Bloch vector decomposition of $|d\rangle \langle d|$
\begin{equation}\label{d-state-decomp}
    |d \rangle \langle d| \;=\; \frac{1}{d} \left(
    -\sqrt{\frac{(d-1)d}{2}} \,\Lambda^{d-1} \,+\, \mathbbm{1} \right) \,,
\end{equation}
where we have applied $\sum_{j=1}^{d-1} |j \rangle \langle j | \,=\, \mathbbm{1} - |d
\rangle
    \langle d|\,$.

Inserting now decomposition \eqref{d-state-decomp} into relation \eqref{recd} we gain the
Bloch vector expansion for $|d-1 \rangle \langle d-1|$ and recurrence relation
\eqref{recl} provides $|d-2 \rangle \langle d-2|$ and so forth. Thus finally we find
\begin{equation} \label{dminusn}
    |d-n \rangle \langle d-n | \;=\; - \sqrt{\frac{d-n-1}{2(d-n)}}
    \,\Lambda^{d-n-1} \,+\, \sum_{k=0}^{n-1} \frac{1}{\sqrt{2(d-n+k+1)(d-n+k)}}
    \,\Lambda^{d-n+k} + \frac{1}{d} \mathbbm{1} ,
\end{equation}
the relation we had to prove, where $d-n=j\,$. $\Box$

\section{The polarization operator basis}

\subsection{Definition and examples}

The polarization operators in the Hilbert-Schmidt space of dimension $d$ are defined as
the following $d \times d$ matrices \cite{varshalovich88, kryszewski06} \,:
\begin{equation} \label{defpo}
    T_{LM} \;=\; \sqrt{\frac{2L+1}{2s+1}} \sum_{k,l =1}^d C^{s m_k}_{s
    m_l , \,LM} \,|k \rangle \langle l | \,.
\end{equation}
The used indices have the properties
\begin{eqnarray}
    & s = \frac{d-1}{2} \,, & \nonumber\\
    & L = 0,1, \ \ldots \ ,2s \,, & \nonumber\\
    & M = -L, -L+1, \ldots, L-1, L \,, & \nonumber\\
    & m_1 = s, \ m_2 = s-1, \ldots ,m_d = -s \,.
\end{eqnarray}
The coefficients $C^{s m_k}_{s m_l , \,LM}$ are identified with the usual Clebsch--Gordan
coefficients $C^{j m}_{j_1 m_1 , \,j_2 m_2}$ of the angular momentum theory and are
displayed explicitly in tables, e.g., in Ref.~\cite{varshalovich88}.\\

For $L=M=0$ the polarization operator is proportional to the identity matrix
\cite{varshalovich88, kryszewski06},
\begin{equation} \label{po00}
    T_{00} \,=\, \frac{1}{\sqrt{d}} \, \mathbbm{1} \,.
\end{equation}
It is shown in Ref.~\cite{kryszewski06} that all polarization operators (except $T_{00}$)
are traceless, in general \emph{not} Hermitian, and that orthogonality relation
\eqref{orthogon} is satisfied
\begin{equation}\label{orthogonal-pob}
    \text{Tr} \, T_{L_1 M_1}^\dag T_{L_2 M_2} \;=\; \delta_{L_1 L_2}
    \delta_{M_1 M_2} \;.
\end{equation}
Therefore the $d^2$ polarization operators \eqref{defpo} form an orthonormal matrix basis
---the polarization operator basis (POB)--- of the Hilbert--Schmidt space of dimension
$d$.\\

\emph{Examples.} The simplest example is of dimension $2$, the qubit. For a qubit the POB
is given by the following matrices ($s = 1/2; L=0,1; M=-1,0,1$)
\begin{eqnarray}\label{pobdim2}
    & T_{00} \;=\; \frac{1}{\sqrt{2}} \left(
    \begin{array}{cc}
        1 & 0 \\
        0 & 1
    \end{array} \right), \quad
    T_{11} \;=\; - \left(
    \begin{array}{cc}
        0 & 1 \\
        0 & 0
    \end{array} \right), & \nonumber\\
    & T_{10} \;=\; \frac{1}{\sqrt{2}} \left(
    \begin{array}{cc}
        1 & 0 \\
        0 & -1
    \end{array} \right), \quad
    T_{1-1} \;=\; \left(
    \begin{array}{cc}
        0 & 0 \\
        1 & 0
    \end{array} \right). &
\end{eqnarray}

For the next higher dimension $d=3$ ($s=1$), the case of qutrits, we get $9$ polarization
operators $T_{LM}$ with $L=0,1,2$ and $M=-L,...,L\,$ and we have
\begin{alignat}{3} \label{pobdim3}
    T_{11} = -\frac{1}{\sqrt{2}}
    &\begin{pmatrix}
        0 & 1 & 0 \\
        0 & 0 & 1 \\
        0 & 0 & 0
    \end{pmatrix}, \quad
    & T_{10} = \frac{1}{\sqrt{2}}
    &\begin{pmatrix}
        1 & 0 & 0 \\
        0 & 0 & 0 \\
        0 & 0 & -1
    \end{pmatrix}, \quad
    & T_{1-1} = \frac{1}{\sqrt{2}}
    &\begin{pmatrix}
        0 & 0 & 0 \\
        1 & 0 & 0 \\
        0 & 1 & 0
    \end{pmatrix}, \nonumber \\
    T_{22} =
    &\begin{pmatrix}
        0 & 0 & 1 \\
        0 & 0 & 0 \\
        0 & 0 & 0
    \end{pmatrix}, \quad
    & T_{21} = \frac{1}{\sqrt{2}}
    &\begin{pmatrix}
        0 & -1 & 0 \\
        0 & 0 & 1 \\
        0 & 0 & 0
    \end{pmatrix}, \quad
    & T_{20} = \frac{1}{\sqrt{6}}
    &\begin{pmatrix}
        1 & 0 & 0 \\
        0 & -2 & 0 \\
        0 & 0 & 1
    \end{pmatrix}, \nonumber \\
    T_{2-1} = \frac{1}{\sqrt{2}}
    &\begin{pmatrix}
        0 & 0 & 0 \\
        1 & 0 & 0 \\
        0 & -1 & 0
    \end{pmatrix}, \quad
    & T_{2-2} =
    &\begin{pmatrix}
        0 & 0 & 0 \\
        0 & 0 & 0 \\
        1 & 0 & 0
    \end{pmatrix}.
\end{alignat}

Then the decomposition of any density matrix into a Bloch vector by using the POB has, in
general, the following form:
\begin{equation} \label{bvpob-d}
    \rho \;=\; \frac{1}{d} \,\mathbbm{1} \,+\, \sum_{L=1}^{2s} \sum_{M=-L}^L b_{LM} T_{LM}
     \;=\; \frac{1}{d} \,\mathbbm{1} \,+\, \vec{b} \cdot\vec{T} \,,
\end{equation}
with the Bloch vector $\vec{b} =
(b_{1-1},b_{10},b_{11},b_{2-2},b_{2-1},b_{20},...,b_{LM})$, where the components are
ordered and given by $b_{LM} = \textnormal{Tr}\,T_{LM}^\dagger \rho\,$. In general the
components $b_{LM}$ are complex since the polarization operators $T_{LM}$ are not
Hermitian. All Bloch vectors lie within a hypersphere of radius $|\vec{b}| \leq
\sqrt{(d-1)/d}\,$.

In $2$ dimensions the Bloch vector $\vec{b} = (b_{1-1},b_{10},b_{11})$ is limited by
$|\vec{b}| \leq \frac{1}{\sqrt{2}}\,$ and forms a spheroid \cite{kryszewski06}, the pure
states occupy the surface and the mixed ones lie in the volume. This decomposition is
fully equivalent to the standard description of Bloch vectors with Pauli matrices.

In higher dimensions, however, the structure of the allowed range of $\vec{b}$ (due to
the positivity requirement $\rho \geq 0$) is quite complicated, as can be seen already
for $d=3$ (for details see Ref.~\cite{kryszewski06}). Nevertheless, pure states are on
the surface, mixed ones lie within the volume and the maximal mixed one corresponds to
$|\vec{b}|=0\,$, thus $|\vec{b}|$ is a kind of measure for the mixedness of a quantum
state.

\subsection{Standard matrix basis expansion by POB}

The standard matrices \eqref{standardmatrices} can be expanded by the POB as
\cite{varshalovich88}
\begin{equation} \label{smpob}
| i \rangle \langle j| \;=\; \sum_L \sum_M \sqrt{\frac{2L+1}{2s+1}} \, C^{s m_i}_{s
m_j,\,LM} \,T_{LM} \,.
\end{equation}
Note that $\sum_M$ is actually fixed by the condition $m_j + M = m_i$.\\

\emph{Proof}. Inserting definition \eqref{defpo} on the right--hand side (RHS) of
equation \eqref{smpob} we find
\begin{eqnarray}
    \textrm{RHS} & \;=\; & \sum_{k,l} \left( \sum_L \frac{2L+1}{2s+1}
        \,C^{sm_i}_{sm_j,\,LM} \,C^{sm_k}_{sm_l,\,LM} \right) |k \rangle
        \langle l| = \nonumber\\
    & \;=\; & \sum_{k,l} \delta_{jl} \,\delta_{ik} \,|k \rangle \langle
        l| \nonumber\\
    & \;=\; & |i \rangle \langle j| \,,
\end{eqnarray}
where we used the sum rule for Clebsch--Gordan coefficients \cite{varshalovich88}
\begin{equation}
    \sum_{c,\gamma} \frac{2c+1}{2b+1} \,C^{b \beta}_{a \alpha,\,c \gamma}
    \,C^{b \beta'}_{a \alpha',\,c \gamma} \;=\; \delta_{\alpha \alpha'}
    \,\delta_{\beta \beta'} \,.
\end{equation}

\section{Weyl operator basis} \label{secwob}

\subsection{Definition and example} \label{secwob-defandexample}

Finally we want to discuss a basis of the Hilbert--Schmidt space of dimension $d$ that
consists of the following $d^2$ operators:
\begin{equation} \label{defwo}
    U_{nm} \;=\; \sum_{k=0}^{d-1} e^{\frac{2 \pi i}{d}\,kn} \,| k \rangle
    \langle (k+m) \,\textrm{mod}\,d| \qquad n,m = 0,1, \ldots ,d-1 \,,
\end{equation}
where we use the standard basis of the Hilbert space.

The operators in notation \eqref{defwo} have been introduced in the context of quantum
teleportation of qudit states \cite{bennett93} and are often called \emph{Weyl operators}
in the literature (see e.g. Refs.~\cite{narnhofer06, baumgartner06, baumgartner07a}). The
$d^2$ operators \eqref{defwo} are unitary and form an orthonormal basis of the
Hilbert--Schmidt space
\begin{equation} \label{orthonormwo}
\text{Tr} \,U_{nm}^{\dag} U_{lj} \;=\; d \,\delta_{nl} \,\delta_{mj}
\end{equation}
(a proof is presented in Appendix~\ref{secproofon}) -- the Weyl operator basis (WOB).
They can be used to create a basis of $d^2$ maximally entangled qudit states
\cite{narnhofer06, werner01, vollbrecht00}.

Clearly the operator $U_{00}$ represents the identity $U_{00} = \mathbbm{1}\,$.\\

\emph{Example.} Let us show the example of dimension $3$, the qutrit case. There the Weyl
operators \eqref{defwo} have the following matrix form
\begin{eqnarray}
    & 
    U_{01} = \begin{pmatrix}
        0 & 1 & 0 \\
        0 & 0 & 1 \\
        1 & 0 & 0
        \end{pmatrix}, \qquad\qquad\quad
    U_{02} = \begin{pmatrix}
        0 & 0 & 1 \\
        1 & 0 & 0 \\
        0 & 1 & 0
        \end{pmatrix}, & \\
    & U_{10} = \begin{pmatrix}
        1 & 0 & 0 \\
        0 & e^{2 \pi i/3} & 0 \\
        0 & 0 & e^{-2 \pi i/3}
        \end{pmatrix}, \quad
    U_{11} = \begin{pmatrix}
        0 & 1 & 0 \\
        0 & 0 & e^{2 \pi i/3} \\
        e^{-2 \pi i/3} & 0 & 0
        \end{pmatrix}, \quad
    U_{12} = \begin{pmatrix}
        0 & 0 & 1 \\
        e^{2 \pi i/3} & 0 & 0 \\
        0 & e^{-2 \pi i/3} & 0
        \end{pmatrix}, & \nonumber\\
    & U_{20} = \begin{pmatrix}
        1 & 0 & 0 \\
        0 & e^{-2 \pi i/3} & 0 \\
        0 & 0 & e^{2 \pi i/3}
        \end{pmatrix}, \quad
    U_{21} = \begin{pmatrix}
        0 & 1 & 0 \\
        0 & 0 & e^{-2 \pi i/3} \\
        e^{2 \pi i/3} & 0 & 0
        \end{pmatrix}, \quad
    U_{22} = \begin{pmatrix}
        0 & 0 & 1 \\
        e^{-2 \pi i/3} & 0 & 0 \\
        0 & e^{2 \pi i/3} & 0
        \end{pmatrix}. & \nonumber
\end{eqnarray}\\

Using the WOB we can decompose quite generally any density matrix into a Bloch vector
\begin{equation} \label{bvwob-d}
    \rho \;=\; \frac{1}{d} \,\mathbbm{1} \,+\, \sum_{n,m=0}^{d-1} b_{nm} U_{nm}
     \;=\; \frac{1}{d} \,\mathbbm{1} \,+\, \vec{b} \cdot \vec{U} \,,
\end{equation}
with $n,m = 0,1, ... ,d-1$ ($b_{00}=0$). The components of the Bloch vector $\vec{b} =
\big(\{b_{nm}\}\big)$ are ordered and given by $b_{nm} =
\textnormal{Tr}\,U_{nm}\,\rho\,$. In general the components $b_{nm}$ are complex since
the Weyl operators are not Hermitian and the complex conjugates fulfil the relation
$b_{n\,m}^\ast = e^{-\frac{2 \pi i}{d}\,nm} \, b_{{-n}{-m}}\,$, which follows easily from
definition \eqref{defwo} together with the hermiticity of $\rho\,$.

All Bloch vectors lie within a hypersphere of radius $|\vec{b}| \leq \sqrt{d-1}/d\,$. For
example, for qutrits the Bloch vector is expressed by
$\vec{b}=(b_{01},b_{02},b_{10},b_{11},b_{12},b_{20},b_{21},b_{22}$) and $|\vec{b}| \leq
\sqrt{2}/3\,$. In $3$ and higher dimensions the allowed range of the Bloch vector is
quite restricted within the hypersphere and the detailed structure is not known yet.

Note that in $2$ dimensions the WOB as well as the GGB coincides with the Pauli matrix
basis and the POB represents a rotated Pauli basis (where $\sigma_{\pm} =
\frac{1}{2}\,(\sigma_1 \pm i\sigma_2)$), in particular
\begin{eqnarray}
\left\{U_{00},U_{01},U_{10},U_{11}\right\} \;&=&\;
\left\{\mathbbm{1},\sigma_1 ,\sigma_3 , i\sigma_2 \right\}
\,,\\
\left\{\mathbbm{1},\lambda^{12}_s,\lambda^{12}_a,\lambda^1 \right\} \;&=&\;
\left\{\mathbbm{1},\sigma_1 ,\sigma_2 , \sigma_3 \right\} \,,\\
\left\{T_{00},T_{11},T_{10},T_{1-1}\right\} \;&=&\;
\left\{\frac{1}{\sqrt{2}}\,\mathbbm{1},\,-\sigma_{+}
,\,\frac{1}{\sqrt{2}}\,\sigma_3 , \,\sigma_{-} \right\} \,.
\end{eqnarray}

\subsection{Standard matrix basis expansion by WOB}

The standard matrices \eqref{standardmatrices} can be expressed by the WOB in the
following way
\begin{equation} \label{smwob}
    |j \rangle \langle k | \;=\; \frac{1}{d}\, \sum_{l=0}^{d-1} e^{-\frac{2 \pi i}{d}
    \,lj} \,U_{l\, (k-j)\,\textrm{mod}\,d} \;.
\end{equation}

\emph{Proof.} We insert the definition of the Weyl operators \eqref{defwo} on the
right--hand side (RHS) of Eq.~\eqref{smwob}, use Eq.~\eqref{complexsumrule} and get
\begin{eqnarray}
    \textrm{RHS} & \;=\; & \frac{1}{d} \,\sum_{l,r=0}^{d-1} e^{\frac{2 \pi i}{d}\,l(r-j)}
    \,|r \rangle \langle (r+k-j)\,\textrm{mod}\,d| \nonumber\\
    & \;=\; & |j \rangle \langle k | \;+\; \frac{1}{d} \sum_{r \neq j,\,r=0}^{d-1}
        \sum_{l=0}^{d-1} e^{\frac{2 \pi i}{d}\,l(r-j)} \,|r \rangle \langle
        (r+k-j)\,\textrm{mod}\,d | \nonumber\\
    & \;=\; & |j \rangle \langle k | \,. \quad\Box
\end{eqnarray}

\section{Isotropic two--qudit state}
\label{seciso}

Now we consider bipartite systems in a $d\times d$ dimensional Hilbert space ${\cal
H}^{\,d}_A \otimes {\cal H}^{\,d}_B$. The observables acting in the subsystems ${\cal
H}_A$ and ${\cal H}_B$ are usually called Alice and Bob in quantum communication.

Quite generally, a density matrix of a two--qudit state acting on ${\cal H}^{\,d}_A
\otimes {\cal H}^{\,d}_B$ can be decomposed in the following way (neglecting the
reference to $A$ and $B$)
\begin{equation} \label{2-quditstategeneral}
    \rho \;=\; \frac{1}{d} \,\mathbbm{1} \otimes \mathbbm{1} \,+\,
    n_i\,\Gamma_i \otimes \mathbbm{1} \,+\, m_i\,\mathbbm{1} \otimes \Gamma_i \,+\,
    c_{ij}\,\Gamma_i \otimes \Gamma_j \,, \qquad
    n_i, m_i, c_{ij} \in \mathbbm{C} \,,
\end{equation}
where $\left\{ \Gamma_i \right\}$ represents some basis in the subspace ${\cal
H}^{\,d}\,$. The term $c_{ij}\,\Gamma_i \otimes \Gamma_j$ always can be diagonalized by
two independent orthogonal transformations on $\Gamma_i$ and $\Gamma_j$ \cite{henley62}.
Altogether there are $(d^{\,2})^2 - 1$ terms.

However, for isotropic two--qudit states ---the case we consider in our paper--- the
second and third term in expression \eqref{2-quditstategeneral} vanish and the fourth
term reduces to $c_{ii}\,\Gamma_i \otimes \Gamma_i$, which implies the vanishing of
$(d^{\,2} - 1)^2 + (d^{\,2} -1) = d^{\,2}(d^{\,2} - 1)$ terms. Consequently, for an
isotropic two--qudit density matrix there remain $d^{\,2} - 1$ independent terms, which
provides the dimension of the corresponding Bloch vector. Thus the isotropic two--qudit
Bloch vector is of the same dimension ---lives in the same subspace--- as the one--qudit
vector, which is a comfortable simplification.

Explicitly, the \emph{isotropic} two--qudit state $\rho_{\alpha}^{(d)}$ is defined as
follows \cite{horodecki99, rains99, horodecki01} \,:
\begin{equation} \label{rhodefiso}
    \rho_{\alpha}^{(d)} \;=\; \alpha \left| \phi_+^d \right\rangle \left\langle \phi_+^d \right|
    \,+\, \frac{1-\alpha}{d^2}\,\mathbbm{1}\,, \quad \alpha \in \mathbbm{R}\,, \quad
    - \frac{1}{d^2-1} \leq \alpha \leq 1 \;,
\end{equation}
where the range of $\alpha$ is determined by the positivity of the state. The state
$\left| \phi^d_+ \right\rangle$, a Bell state, is maximally entangled and given by
\begin{equation} \label{defmaxent}
    \left| \phi^d_+ \right\rangle \;=\; \frac{1}{\sqrt{d}} \,
    \sum_j \left| j \right\rangle \otimes \left| j \right\rangle\;,
\end{equation}
where $\left\{ \left| j \right\rangle \right\}$ denotes the standard basis of the
d--dimensional Hilbert space.

\subsection{Expansion into GGB}

Let us first calculate the Bloch vector notation for the Bell state $\left| \phi_+^d
\right\rangle \left\langle \phi_+^d \right|$ in the GGB. It is convenient to split the
state into two parts
\begin{eqnarray}\label{maxentstandbasis}
    \left| \phi_+^d \right\rangle \left\langle \phi_+^d
        \right| & \;=\; & \frac{1}{d} \sum_{j,k =1}^{d} |j \rangle \langle k |
        \otimes |j \rangle \langle k| \nonumber\\
    & \;=\; & A \,+\, B \,,
\end{eqnarray}
where $A$ and $B$ are defined by
\begin{eqnarray}\label{isoggmA}
    A \;&:=&\; \frac{1}{d} \sum_{j < k} |j \rangle \langle k| \otimes | j
    \rangle \langle k | \,+\, \frac{1}{d} \sum_{j < k} |k \rangle \langle j| \otimes | k
    \rangle \langle j | \,,\\
    B \;&:=&\; \frac{1}{d} \sum_{j} |j \rangle \langle j| \otimes |
    j \rangle \langle j | \,,\label{isoggmB}
\end{eqnarray}
and to calculate the two terms separately.

For term $A$ we use the standard matrix expansion \eqref{smggb} for the case $j \neq k$
and get
\begin{eqnarray}
    A & \;=\; & \frac{1}{4d} \left[ \sum_{j<k} \left( \Lambda^{jk}_s + i
        \Lambda^{jk}_a \right) \otimes \left( \Lambda^{jk}_s + i
        \Lambda^{jk}_a \right) \,+\, \sum_{j<k} \left( \Lambda^{jk}_s - i
        \Lambda^{jk}_a \right) \otimes \left( \Lambda^{jk}_s - i
        \Lambda^{jk}_a \right) \right] \nonumber\\
    & \;=\; & \frac{1}{2d} \,\sum_{i<j} \left( \Lambda^{jk}_s \otimes
    \Lambda^{jk}_s \,-\, \Lambda^{jk}_a \otimes
    \Lambda^{jk}_a \right) \,.
\end{eqnarray}
For term $B$ we need the case $j = k$ in expansion \eqref{smggb} and obtain after some
calculations (the details are presented in Appendix \ref{termB})
\begin{eqnarray}\label{ggb-term-B}
    B \;\;=\;\; \frac{1}{2d} \,
    \sum_{m=1}^{d-1} \Lambda^m \otimes \Lambda^m \,+\, \frac{1}{d^2} \,\mathbbm{1} \otimes
    \mathbbm{1} \,.
\end{eqnarray}
Thus all together we find the following GGB Bloch vector notations, for the Bell state
\eqref{maxentstandbasis}
\begin{eqnarray}\label{maxentggb}
    \left| \phi_+^d \right\rangle \left\langle \phi_+^d \right|  \;\;=\;\;
    \frac{1}{d^2} \,\mathbbm{1} \otimes \mathbbm{1} \,+\, \frac{1}{2d} \;\Lambda \,,
\end{eqnarray}
and for the isotropic two--qudit state \eqref{rhodefiso}
\begin{eqnarray}\label{isoggb}
    \rho^{(d)}_\alpha \;\;=\;\; \frac{1}{d^2} \,\mathbbm{1} \otimes
    \mathbbm{1} \,+\, \frac{\alpha}{2d} \;\Lambda \,,
\end{eqnarray}
where we defined
\begin{eqnarray}\label{Bloch-Lambda}
    \Lambda \;:=\;\; \sum_{i<j} \Lambda^{jk}_s \otimes
    \Lambda^{jk}_s \,-\,  \sum_{i<j} \Lambda^{jk}_a \otimes
    \Lambda^{jk}_a \,+\, \sum_{m=1}^{d-1} \Lambda^m \otimes \Lambda^m \,.
\end{eqnarray}

\subsection{Expansion into POB}

Now we calculate the Bell state $\left| \phi_+^d \right\rangle \left\langle \phi_+^d
\right|$ in the POB. Using expansion \eqref{smpob} and the sum rule for the
Clebsch--Gordan coefficients \cite{varshalovich88}
\begin{equation}\label{CGsumrule}
    \sum_{\alpha, \gamma} C^{c \gamma}_{a \alpha , b \beta} \,C^{c \gamma}_{a \alpha ,
    b' \beta '} \;=\; \frac{2c+1}{2b+1} \,\,\delta_{b b'} \,\delta_{\beta \beta '} \,,
\end{equation}
we obtain
\begin{eqnarray} \label{maxentpob}
    \left| \phi_+^d \right\rangle \left\langle \phi_+^d
        \right| & \;=\; & \frac{1}{d} \sum_{i,j =1}^{d} |i \rangle \langle j |
        \otimes |i \rangle \langle j| \nonumber\\
    & \;=\; & \frac{1}{d} \sum_{L,L'} \frac{\sqrt{(2L+1)(2L'+1)}}{2s+1} \left(
        \sum_{i,j} C^{s m_i}_{s m_j , LM} C^{s m_i}_{s m_j, L' M}
        \right) T_{LM} \otimes T_{L'M} \nonumber\\
    & \;=\; & \frac{1}{d} \sum_{L,L'} \frac{\sqrt{(2L+1)(2L'+1)}}{2L+1}
        \,\, \delta_{L,L'} \, T_{LM} \otimes T_{L' M} \nonumber\\
    & \;=\; & \frac{1}{d} \sum_{L} T_{LM} \otimes T_{LM} \nonumber\\
    & \;=\; & \frac{1}{d^2} \,\mathbbm{1} \otimes \mathbbm{1} \,+\,
    \frac{1}{d} \, T \,,
\end{eqnarray}
where we extracted the unity (recall Eq.~\eqref{po00}) and defined
\begin{equation}\label{Bloch-T}
    T \;:=\; \sum_{L,M \neq 0,0} T_{LM} \otimes T_{LM} \,.
\end{equation}
Result \eqref{maxentpob} provides the POB Bloch vector notation of the isotropic
two--qudit state \eqref{rhodefiso}
\begin{equation}\label{isopob}
    \rho^{(d)}_\alpha \;=\; \frac{1}{d^2} \,\mathbbm{1} \otimes
    \mathbbm{1} \,+\, \frac{\alpha}{d} \, T \,.
\end{equation}

\subsection{Expansion into WOB}

Finally we present the Bell state in the WOB (the details for our approach using the
standard matrix expression \eqref{smwob} can be found in the Appendix \ref{BellWOB}, see
also Ref.~\cite{narnhofer06})
\begin{equation} \label{maxentwob2}
    \left| \phi_+^d \right\rangle \left\langle \phi_+^d \right| \;=\;
    \frac{1}{d^2} \,\mathbbm{1} \otimes \mathbbm{1} \,+\, \frac{1}{d^2} \, U \,,
\end{equation}
with
\begin{equation} \label{defu}
    U \;:=\; \sum_{l,m = 0}^{d-1} U_{lm} \otimes
        U_{-lm} \,, \qquad (l,m) \neq (0,0) \,,
\end{equation}
where negative values of the index $l$ have to be considered as $mod \ d\,$, and from
formula \eqref{maxentwob2} we find the WOB Bloch vector notation of the isotropic
two--qudit state
\begin{equation} \label{isowob}
    \rho^{(d)}_\alpha \;=\; \frac{1}{d^2} \,\mathbbm{1} \otimes \mathbbm{1}
    \,+\, \frac{\alpha}{d^2} \, U \,.
\end{equation}

\section{Applications of the matrix bases}\label{secapp}

\subsection{Entangled isotropic two--qudit states}\label{secqudits}

In Ref.~\cite{bertlmann05} the connection between the Hilbert--Schmidt (HS) measure of
entanglement \cite{witte99,ozawa00,bertlmann02} and the optimal entanglement witness is
investigated. Explicit calculations for both quantities are presented in case of
isotropic qutrit states. For higher dimensions, the isotropic two--qudit states, the
above quantities are determined as well but in terms of a rather general matrix basis
decomposition. With the results of the present paper we can calculate all quantities
explicitly. Let us recall the basic notations we need.

The HS \emph{measure} is defined as the minimal HS distance of an
entangled state $\rho_{\rm{ent}}$ to the set of separable states $S$
\begin{equation} \label{defhs}
    D(\rho_{\rm{ent}}) \;:=\; \min_{\rho \in S} \left\| \rho - \rho_{\rm{ent}} \right\|
    \;=\; \left\| \rho_0 - \rho_{\rm{ent}} \right\| \,,
\end{equation}
where $\rho_0$ denotes the nearest separable state, the minimum of the HS distance.

An \emph{entanglement witness} $A \in {\cal A}$ (${\cal A} = {\cal A}_A \otimes {\cal
A}_B\,$, the HS space of operators acting on the Hilbert space of states) is a Hermitian
operator that ``detects'' the entanglement of a state $\rho_{\rm ent}$ via inequalities
\cite{horodecki96, terhal00, terhal02, bertlmann02}.

\begin{defi}
An entanglement witness $A$ is a Hermitian operator with the following properties:
The expectation value of $A$ is negative for an entangled
state, whereas it is non--negative for any separable state.
\begin{eqnarray} \label{defentwit}
    \left\langle \rho_{\rm ent},A \right\rangle \;=\; \textnormal{Tr}\, \rho_{\rm ent} A
    & \;<\; & 0 \,,\nonumber\\
    \left\langle \rho,A \right\rangle = \textnormal{Tr}\, \rho A & \;\geq\; & 0 \qquad
    \forall \rho \in S \,.
\end{eqnarray}
\end{defi}

The fact, however that there exists an operator satisfying inequalities \eqref{defentwit}
for \emph{any} entangled state, i.e. that the definition is meaningful, has to be proved;
it follows from the Hahn--Banach Theorem of functional analysis (for a simple geometric
approach, see Ref.~\cite{bertlmann05}).\\

\begin{figure}
    \centering
        \includegraphics[width=0.40\textwidth]{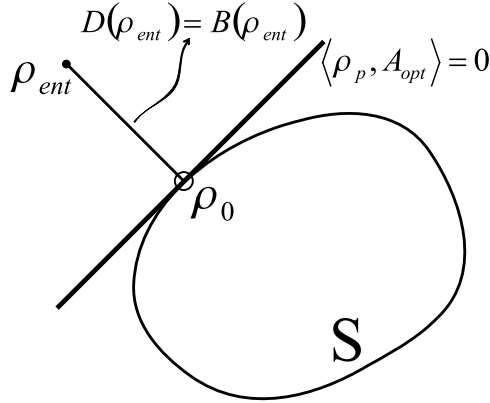}
    \caption{Illustration of the Bertlmann--Narnhofer--Thirring Theorem~\eqref{bnttheorem}}
    \label{figbnttheo}
\end{figure}

An entanglement witness is ``optimal'', denoted by $A_{\rm{opt}}\,$, if apart from
Eq.~(\ref{defentwit}) there exists a separable state $\rho_0 \in S$ such that
\begin{equation}
    \left\langle \rho_0 ,A_{\rm{opt}} \right\rangle \;=\; 0 \,.
\end{equation}
The operator $A_{\rm{opt}}$ defines a tangent plane to the set of separable states $S$
and all states $\rho_p$ with $\left\langle \rho_p ,A_{\rm{opt}} \right\rangle \;=\; 0 \,$
lie within that plane; see Fig.~\ref{figbnttheo}.
\\

Let us call the lower one of the inequalities \eqref{defentwit} an \textit{entanglement
witness inequality}, short EWI. It detects entanglement whereas a Bell inequality
determines non--locality. Rewriting Eq.~\eqref{defentwit} as
\begin{equation} \label{gbi}
    \left\langle \rho,A \right\rangle \,-\, \left\langle \rho_{\rm ent},A \right\rangle
    \;\geq\; 0 \qquad \forall \rho \in S \,,
\end{equation}
the \emph{maximal violation} of the EWI is defined by
\begin{equation} \label{maxviolationgbi}
    B(\rho_{\rm ent}) \;=\; \max_{A, \, \left\| A - a \mathbbm{1} \right\| \leq 1}
    \left( \min_{\rho \in S} \left\langle \rho,A \right\rangle \,-\,
    \left\langle \rho_{\rm ent},A \right\rangle \right),
\end{equation}
where the maximum is taken over all possible entanglement witnesses $A$, suitably
normalized.\\

\vspace{0.3cm}

Then an interesting connection between the HS measure and the concept of entanglement
witnesses is given by the Bertlmann--Narnhofer--Thirring Theorem, illustrated in
Fig.~\ref{figbnttheo} \cite{bertlmann02}.

\begin{thm}\label{BNT-theorem}
\hspace{2cm}
\begin{enumerate}
    \item[i)] The maximal violation of the EWI is equal to the minimal distance of
    $\rho_{\rm ent}$ to the set $S$
    \begin{equation} \label{bnttheorem}
    B(\rho_{\rm ent}) \;=\; D(\rho_{\rm ent}) \;.
    \end{equation}
    \item[ii)] The maximal violation of the EWI is attained for an optimal entanglement
    witness
    \begin{equation} \label{entwitmaxviolation}
    A_{\rm{opt}} \;=\; \frac{\rho_0 - \rho_{\rm ent} \,-\, \left\langle \rho_0 ,
    \rho_0 - \rho_{\rm ent} \right\rangle \mathbbm{1}}{\left\| \rho_0 - \rho_{\rm ent}
    \right\|}\;.
    \end{equation}
\end{enumerate}
\end{thm}
Thus the calculation of the optimal entanglement witness $A_{\rm{opt}}$ to a given
entangled state $\rho_{\rm ent}$ reduces to the determination of the nearest separable
state $\rho_0\,$. In special cases $\rho_0$ is detectable but in general its detection is
quite a difficult task. We are able to find the nearest separable state by working with
Lemma~\ref{lemsepablenearest}, a method we call \emph{guess method} \cite{bertlmann05}.
\begin{lem} \label{lemsepablenearest}
    A state $\tilde{\rho}$ is equal to the nearest separable state $\rho_0$
    if and only if the operator
    \begin{equation} \label{ctilde}
    \tilde{C} \;=\; \frac{\tilde{\rho} - \rho_{\rm ent} \,-\, \left\langle \tilde{\rho} ,
    \tilde{\rho} - \rho_{\rm ent} \right\rangle \mathbbm{1}}{\left\| \tilde{\rho} -
    \rho_{\rm ent} \right\|}
\end{equation}
is an entanglement witness.
\end{lem}
Lemma~\ref{lemsepablenearest} probes if a guess $\tilde{\rho}$ is indeed correct for the
nearest separable state. Then operator $\tilde{C}$ represents the optimal entanglement
witness $A_{\rm{opt}}$ \eqref{entwitmaxviolation}.\\

Now let us apply the matrix bases we discussed in the previous sections and calculate the
quantities introduced above. As an entangled state we consider the isotropic two--qudit
state $\rho_\alpha^{(d), \, \rm{ent}}$, that is the state $\rho_{\alpha}^{(d)}$
\eqref{rhodefiso} for $\frac{1}{d+1} < \alpha \leq 1$.

Starting with the GGB we can express that state in our Bloch vector notation by formula
\eqref{isoggb}. By using Lemma~\ref{lemsepablenearest} we find that the nearest separable
state is reached at $\alpha = \frac{1}{d+1}$
\begin{equation} \label{quditisoresult1}
    \rho^{(d)}_0 \;=\; \rho_{\alpha = \frac{1}{d+1}}^{(d)} \;=\;
    \frac{1}{d^2} \,\mathbbm{1} \otimes \mathbbm{1} \,+\,
    \frac{1}{2 \, d(d+1)} \, \Lambda \,.
\end{equation}
It provides the HS measure
\begin{equation} \label{quditisoresult2}
    D(\rho_{\alpha , \,\rm{ent}}^{(d)}) \;=\; \left\| \rho_0^{(d)} -
        \rho_{\alpha , \,\rm{ent}}^{(d)} \right\| \;=\; \frac{\sqrt{d^2-1}}{d}\,
        \left( \alpha \,-\, \frac{1}{d+1} \right) \,,
\end{equation}
and the optimal entanglement witness \eqref{entwitmaxviolation}
\begin{equation} \label{quditisoresult3}
    A_{\rm{opt}}(\rho_{\alpha , \,\rm{ent}}^{(d)}) \;=\; \frac{1}{d} \,\sqrt{\frac{d-1}{d+1}}\,
    \mathbbm{1} \otimes \mathbbm{1} \,-\, \frac{1}{2\sqrt{d^2-1}} \; \Lambda \,,
\end{equation}
where we used the HS norm $\| \Lambda \| = 2 \sqrt{d^2-1}\,$.

Clearly, the maximal violation $B$ of the EWI equals the HS measure $D$
\begin{eqnarray}\label{BequalsD}
    B(\rho_{\alpha , \,\rm{ent}}^{(d)}) &\;=\;& - \left\langle \rho_{\alpha ,
    \,\rm{ent}}^{(d)} , A_{\rm{opt}} \right\rangle \nonumber\\
    &\;=\;& \frac{\sqrt{d^2-1}}{d} \left( \alpha \,-\, \frac{1}{d+1} \right) \;=\;
    D(\rho_{\alpha , \,\rm{ent}}^{(d)}) \,.
\end{eqnarray}

For expressing above quantities by the matrix bases POB and WOB it suffices to calculate
the proportionality factors between $\Lambda$, $T$ and $U\,$. By comparison of the three
forms for the isotropic qudit state \eqref{isoggb}, \eqref{isopob} and \eqref{isowob} we
find
\begin{equation}
    \Lambda \;=\; 2 \, T \qquad  \textrm{and} \qquad T \;=\; \frac{1}{d} \, U \,.
\end{equation}
It provides the following expressions, for the POB
\begin{equation}
    \rho^{(d)}_0 \;=\; \rho_{\alpha = \frac{1}{d+1}}^{(d)} \;=\;
    \frac{1}{d^2} \,\mathbbm{1} \otimes \mathbbm{1} \,+\,
    \frac{1}{d(d+1)} \; T \,,
\end{equation}
\begin{equation}
    A_{\rm{opt}}(\rho_{\alpha , \,\rm{ent}}^{(d)}) \;=\; \frac{1}{d} \,\sqrt{\frac{d-1}{d+1}}
    \,\mathbbm{1} \otimes \mathbbm{1} \,-\, \frac{1}{\sqrt{d^2-1}} \; T \,,
\end{equation}
and for the WOB
\begin{equation}
    \rho^{(d)}_0 \;=\; \rho_{\alpha = \frac{1}{d+1}}^{(d)} \;=\;
    \frac{1}{d^2} \,\mathbbm{1} \otimes \mathbbm{1} \,+\,
    \frac{1}{d^2(d+1)} \; U \,,
\end{equation}
\begin{equation}
    A_{\rm{opt}}(\rho_{\alpha , \,\rm{ent}}^{(d)}) \;=\; \frac{1}{d} \,\sqrt{\frac{d-1}{d+1}}
    \,\mathbbm{1} \otimes \mathbbm{1} \,-\, \frac{1}{d\sqrt{d^2-1}} \; U \,.
\end{equation}
Of course, the HS measure $D(\rho_{\alpha , \,\rm{ent}}^{(d)})$ remains the same
expression \eqref{quditisoresult2} independent of the chosen matrix basis, which can
easily be verified using $\| T \| = \sqrt{d^2-1}$ and $\| U \| = d \sqrt{d^2-1}\,$.

\subsection{Entanglement witness representation for experiments}\label{witness-experimental}

Entanglement witnesses are Hermitian operators and therefore observables that should be
measurable in a given experimental set--up and thus provide an experimental verification
of entanglement. The quantity to be measured is the expectation value
\begin{equation}
    \left\langle A \right\rangle \:=\,\textnormal{Tr} A \rho
\end{equation}
of an entanglement witness $A$ for some state $\rho$. If $\left\langle A \right\rangle <
0$ then the state $\rho$ is entangled. But which measurements have to be performed?

Obviously it is appropriate to express the entanglement witness in terms of generalized
Gell-Mann matrices \eqref{ggms}--\eqref{ggmd}, since they are Hermitian. For $d=3$
---qutrits--- the Gell-Mann matrices \eqref{defgellmann-s}--\eqref{defgellmann-d} can be
expressed in terms of eight ``physical'' operators, the observables $S_x$, $S_y$, $S_z$,
$S_x^2$, $S_y^2$, $\{S_x,S_y\}$, $\{S_y,S_z\}$, $\{S_z,S_x\}$ of a spin--1 system, where
$\vec{S}= (S_x,S_y,S_z)$ is the spin operator and $\{S_i,S_j\} = S_iS_j + S_jS_i$ (with
$i,j = x,y,z$) denotes the corresponding anticommutator. The decomposition of the
Gell-Mann matrices into spin--1 operators is as follows (for a similar expansion, see
Ref.~\cite{mendas06}):
\begin{alignat}{2} \label{gellmann-spin1}
    \lambda^{12}_s &\,=\, \frac{1}{\sqrt{2} \hbar^2} \left( \hbar S_x +
        \{S_z,S_x\} \right), \qquad
        & \lambda^{13}_s &\,=\, \frac{1}{\hbar^2}
        \left( S_x^2 - S_y^2 \right), \nonumber\\
    \lambda^{23}_s &\,=\, \frac{1}{\sqrt{2} \hbar^2}
        \left( \hbar S_x - \{S_z,S_x \} \right), \qquad
        & \lambda^{12}_a &\,=\, \frac{1}{\sqrt{2} \hbar^2}
        \left( \hbar S_y + \{S_y,S_z \} \right), \nonumber\\
    \lambda^{13}_a &\,=\, \frac{1}{\hbar^2}
        \left\{ S_x, S_y \right\}, \qquad
        & \lambda^{23}_a &\,=\, \frac{1}{\sqrt{2} \hbar^2}
        \left( \hbar S_y - \{S_y,S_z \} \right), \nonumber\\
    \lambda^1 &\,=\, 2 \, \mathbbm{1} + \frac{1}{2 \hbar^2} \left( \hbar
        S_z - 3 S_x^2 -3 S_y^2 \right), \qquad
        & \lambda^2 &\,=\, \frac{1}{\sqrt{3}} \left( - 2 \, \mathbbm{1} +
        \frac{3}{2 \hbar^2} \left( \hbar S_z + S_x^2 + S_y^2 \right)
        \right).
\end{alignat}
All operators can be represented by the following matrices:
\begin{align}\label{spin-1-matrices}
    &S_x \;=\; \frac{\hbar}{\sqrt{2}}
    \begin{pmatrix}
        0 & 1 & 0 \\
        1 & 0 & 1 \\
        0 & 1 & 0
    \end{pmatrix}, \quad
    S_y \;=\; \frac{\hbar}{\sqrt{2}}
    \begin{pmatrix}
        0 & -i & 0 \\
        i & 0 & -i \\
        0 & i & 0
    \end{pmatrix}, \quad
    S_z \;=\; \hbar
    \begin{pmatrix}
        1 & 0 & 0 \\
        0 & 0 & 0 \\
        0 & 0 & -1
    \end{pmatrix}, \nonumber\\
    &S_x^2 \;=\; \,\frac{\hbar^2}{2}\,
    \begin{pmatrix}
        1 & 0 & 1 \\
        0 & 2 & 0 \\
        1 & 0 & 1
    \end{pmatrix}, \quad
    S_y^2 \;=\; \,\frac{\hbar^2}{2}\,
    \begin{pmatrix}
        1 & 0 & -1 \\
        0 & 2 & 0 \\
        -1 & 0 & 1
    \end{pmatrix}, \quad & \nonumber\\
    &\{S_x,S_y\} \;=\; \;\hbar^2\,
    \begin{pmatrix}
        0 & 0 & -i \\
        0 & 0 & 0 \\
        i & 0 & 0
    \end{pmatrix}, \quad
    \{S_y,S_z\} \;=\; \frac{\hbar^2}{\sqrt{2}}
    \begin{pmatrix}
        0 & -i & 0 \\
        i & 0 & i \\
        0 & -i & 0
    \end{pmatrix}, \quad \nonumber\\
    &\{S_z,S_x\} \;=\; \frac{\hbar^2}{\sqrt{2}}
    \begin{pmatrix}
        0 & 1 & 0 \\
        1 & 0 & -1 \\
        0 & -1 & 0
    \end{pmatrix} .
\end{align}
Thus we can express any observable on a $n$--qutrit Hilbert space ---a composite system
of $n$ particles with $3$ degrees of freedom--- in terms of above spin operators
\eqref{spin-1-matrices}.

As an example we want to study the entanglement witness for the isotropic two--qutrit
state, i.e. state \eqref{rhodefiso} for $d=3$. In this case we obtain for the optimal
entanglement witness
\begin{equation}
    A_{\rm iso} \;=\; \frac{1}{3\sqrt{2}} \,\left( \mathbbm{1} \otimes \mathbbm{1}
    \,-\, \frac{3}{4} \,\Lambda \right) \,,
\end{equation}
(i.e. Eq.~\eqref{quditisoresult3} for $d=3$) where the operator $\Lambda$ is defined in
Eq.~\eqref{Bloch-Lambda}.

Expressing the Gell-Mann matrices in $\Lambda$ \eqref{Bloch-Lambda} by the spin operator
decomposition \eqref{gellmann-spin1} we find for the expectation value of the
entanglement witness $A_{\rm iso}$
\begin{equation}
    \left\langle A_{\rm iso} \right\rangle \;=\; \frac{1}{3\sqrt{2}}
    \,\langle \mathbbm{1} \otimes \mathbbm{1} \rangle  \,-\,
    \frac{1}{4\sqrt{2}} \,\left\langle \Lambda \right\rangle \,,
\end{equation}
where
\begin{align}\label{expectvalue-Lambda}
    \left\langle \Lambda \right\rangle \,&=\, \frac{1}{\hbar^2} \Big(
        \langle S_x \otimes S_x \rangle \,-\, \langle S_y \otimes S_y
        \rangle \,+\, \langle S_z \otimes S_z \rangle \Big) \,+\,
        \frac{16}{3} \,\langle \mathbbm{1} \otimes \mathbbm{1} \rangle \nonumber\\
    &-\, \frac{4}{\hbar^2} \Big( \langle \mathbbm{1} \otimes S_x^2
        \rangle \,+\, \langle \mathbbm{1} \otimes S_y^2 \rangle \,+\, \langle
        S_x^2 \otimes \mathbbm{1} \rangle \,+\, \langle S_y^2 \otimes
        \mathbbm{1} \rangle \Big) \nonumber\\
    &+\, \frac{4}{\hbar^4} \Big( \langle S_x^2 \otimes S_x^2 \rangle
        \,+\, \langle S_y^2 \otimes S_y^2 \rangle \Big) \,+\,
        \frac{2}{\hbar^4} \Big( \langle S_x^2 \otimes S_y^2 \rangle
        \,+\, \langle S_y^2 \otimes S_x^2 \rangle \Big) \nonumber\\
    &+\, \frac{1}{\hbar^4} \Big( \langle \{ S_z,S_x \} \otimes \{ S_z,S_x
        \} \rangle \,-\, \langle \{ S_y,S_z \} \otimes \{ S_y,S_z
        \} \rangle \,-\, \langle \{ S_x,S_y \} \otimes \{ S_x,S_y
        \} \rangle \Big) \,.
\end{align}
Decomposition \eqref{expectvalue-Lambda} has to be determined experimentally by measuring
the several expectation values with the set--ups on both Alice's and Bob's side.

The advantage of the entanglement witness procedure is that for an experimental outcome
$\left\langle A_{\rm iso} \right\rangle < 0$ the considered quantum state is definitely
entangled, whereas in case of Bell inequalities a violation detects \textit{nonlocal}
states. That means by the entanglement witness procedure we are able to detect more
entangled states than with Bell inequalities. The amount of measurement steps necessary
to determine an entanglement witness is about the same as in the Bell inequality
procedure (see, e.g., Refs.~\cite{collins02, kaszlikowski02, groeblacher06, vaziri02}).

\section{Conclusion}\label{conclusion}

In this paper we present three different matrix bases which are quite useful to decompose
density matrices for higher dimensional qudits. These are the generalized Gell-Mann
matrix basis, the polarization operator basis and the Weyl operator basis. Each
decomposition we identify with a vector, the so-called Bloch vector.

Considering just one--particle states we observe the following features: The generalized
Gell-Mann matrix basis is easy to construct, the matrices correspond to the standard
SU(N) generators ($N=d$), but in general (in $d$ dimensions) it is rather unpractical to
work with the diagonal matrices \eqref{ggmd} due to their more complicated definition. On
the other hand, the Bloch vector itself has real components, which is advantageous, they
can be expressed as expectation values of measurable quantities. For example, in $3$
dimensions the Gell-Mann matrices are Hermitian and the Bloch vector components can be
expressed by expectation values of spin 1 operators. The polarization operator basis is
also easy to set up, all you need to know are the Clebsch--Gordan coefficients which you
find tabulated in the literature. However, the Bloch vector contains complex components.
For the Weyl operator basis the corresponding operators are again simple to construct,
they are non--Hermitian but unitary. The Bloch vector itself has a very simple structure,
however, with complex components. Let us note that in $2$ dimensions all bases are
equivalent since they correspond to Pauli matrices or linear combinations thereof.

In case of two--qudits we have studied the isotropic states explicitly and find the
following: In the generalized Gell-Mann matrix basis the Bloch vector \eqref{isoggb} with
expression \eqref{Bloch-Lambda} is more complicated to construct, in particular the
diagonal part $B$ \eqref{ggb-term-B} (see Appendix \ref{termB}). In the polarization
operator basis the Bloch vector \eqref{isopob} with expression \eqref{Bloch-T} can be
easily set up by the knowledge of the Clebsch--Gordon coefficient sum rule
\eqref{CGsumrule} and in the Weyl operator basis the Bloch vector \eqref{isowob} with
definition \eqref{defu} is actually most easily to construct.

The Hilbert--Schmidt measure of entanglement can be calculated explicitly for all
isotropic two--qudit states and we want to emphasize its interesting connection to the
maximal violation of the entanglement witness inequality, Theorem~\ref{BNT-theorem}.

For the experimental realization of an entanglement witness the generalized Gell-Mann
matrix basis is the appropriate one since the generalized Gell-Mann matrices are
Hermitian. For a different task, however, the determination of the geometry of
entanglement the Weyl operator basis turns out to be optimal. In our example of the
entangled isotropic two--qutrit state the entanglement witness can be expressed by
experimental quantities, the expectation values of spin--1 measurements. In this way one
can experimentally find out whether a state is entangled or not, i.e., we can obtain
rather precise information on the quality of entanglement.

Quite generally, the Bloch vector decomposition into one of the three matrix bases is of
particular advantage in the construction of entanglement witnesses. It turns out that if
the coefficients of the decomposition satisfy a certain condition the considered operator
represents an entanglement witness, i.e. satisfies inequalities \eqref{defentwit}
(for details see Ref.~\cite{bertlmann08}).

\begin{acknowledgments}

We would like to thank Beatrix Hiesmayr and Heide Narnhofer for helpful discussions. This
research has been financially supported by FWF project CoQuS No W1210-N16 of the
Austrian Science Foundation and by F140-N Research Grant of the University of Vienna.

\end{acknowledgments}

\appendix

\section{}

\subsection{Proof of Orthogonality of GGB}\label{orthogonality}

We want to proof condition \eqref{orthogon} for the GGB which consists of the $d^2-1$ GGM
\eqref{ggms}, \eqref{ggma}, \eqref{ggmd} and the $d \times d$ unity $\mathbbm{1}$. Since
all GGM are Hermitian (thus $\textnormal{Tr} A_i^{\dag} A_j = \textnormal{Tr} A_i A_j =
\textnormal{Tr} A_j A_i$) it suffices to proof the following conditions:
\begin{eqnarray}
    \textnormal{Tr} \,\Lambda^{jk}_s \Lambda^{mn}_s & \;=\; & 2 \,\delta^{jm}
        \delta^{kn} \label{proofprop5}\\
    \textnormal{Tr} \,\Lambda^{jk}_a \Lambda^{mn}_a & \;=\; & 2 \,\delta^{jm}
        \delta^{kn} \label{proofprop6}\\
    \textnormal{Tr} \,\Lambda^{l} \Lambda^{m} & \;=\; & 2 \,\delta^{lm}
        \label{proofprop7} \\
    \textnormal{Tr} \,\Lambda^{jk}_a \Lambda^{mn}_s & \;=\; & 0 \label{proofprop8} \\
    \textnormal{Tr} \,\Lambda^{jk}_s \Lambda^{m} & \;=\; & 0 \label{proofprop9} \\
    \textnormal{Tr} \,\Lambda^{jk}_a \Lambda^{m} & \;=\; & 0 \,. \label{proofprop10}
\end{eqnarray}

\emph{Proof of condition \eqref{proofprop5}.} Inserting definition \eqref{ggms} we have
\begin{eqnarray} \label{proofss}
    \textnormal{Tr} \,\Lambda^{jk}_s \Lambda^{mn}_s & \;=\; &
        \sum_{l=1}^{d} \langle l | \left( |j \rangle \langle k | \,+\, |
        k  \rangle \langle j | \right) \left( |m \rangle \langle n | \,+\, |
        n  \rangle \langle m | \right) | l \rangle \nonumber\\
    & \;=\; & \sum_l \left(
    \langle l | j \rangle \langle k | m \rangle \langle n | l \rangle \,+\,
    \langle l | j \rangle \langle k | n \rangle \langle m | l \rangle \,+\,
    \langle l | k \rangle \langle j | m \rangle \langle n | l \rangle \,+\,
    \langle l | k \rangle \langle j | n \rangle \langle m | l
    \rangle \right) \nonumber\\
    & \;=\; & \delta^{jn} \delta^{km} \,+\, \delta^{jm} \delta^{kn} \,+\,
        \delta^{kn} \delta^{jm} \,+\, \delta^{km} \delta^{jn} \nonumber\\
    & \;=\; & 2 \,\delta^{jm} \delta^{kn} \,,
\end{eqnarray}
where we used in the last step that $\delta^{jn} \delta^{km} = 0$ since we have $j<k$
\emph{and} $m<n$.\\

\emph{Proof of condition \eqref{proofprop6}.} This case is equivalent to the one before
apart from changed signs that do not matter
\begin{eqnarray} \label{proofaa}
    \textnormal{Tr} \,\Lambda^{jk}_a \Lambda^{mn}_a & \;=\; &
        -\,\delta^{jn} \delta^{km} \,+\, \delta^{jm} \delta^{kn} \,+\,
        \delta^{kn} \delta^{jm} \,-\, \delta^{km} \delta^{jn} \nonumber\\
    & \;=\; & 2 \,\delta^{jm} \delta^{kn} \,.
\end{eqnarray}

\emph{Proof of condition \eqref{proofprop7}.} Using definition \eqref{ggmd} and denoting
\begin{equation} \label{defcdiagggm}
    C_l = \sqrt{\frac{2}{l(l+1)}} \,,
\end{equation}
where $l \leq m$ without loss of generality, we get
\begin{eqnarray}
    \textnormal{Tr} \,\Lambda^l \Lambda^m & \;=\; & C_l C_m \sum_{p=1}^d
        \Big( \sum_{k=1}^l \sum_{n=1}^m \langle p | k \rangle
        \langle k | n \rangle \langle n | p \rangle \,+\, l m \langle p
        | l+1 \rangle \langle l+1 | m+1 \rangle \langle m+1 | p
        \rangle \nonumber\\
    & & - \,m \sum_{k=1}^l \langle p | k \rangle \langle k | m+1
        \rangle \langle m+1 | p \rangle \,-\, l \sum_{n=1}^m \langle p |
        l+1 \rangle \langle l+1 | n \rangle \langle n | p \rangle
        \Big)
        \nonumber\\
    & \;=\; & C_l C_m \left( l \,+\, l m \, \delta^{lm} \,-\, m \sum_{k=1}^{l}
    \delta ^{k(m+1)} \,-\, l \sum_{n=1}^{m} \delta ^{n(l+1)} \right) \,.
\end{eqnarray}
Using the fact that $\delta^{k (m+1)} = 0$ for $m \geq k$ and
\begin{equation}
    l \sum_{n=1}^m \delta^{n (l+1)} \;=\; \begin{cases}
        0 & \text{if} \ l = m \\
        l & \text{if} \ l < m
        \end{cases}
\end{equation}
we obtain
\begin{equation}
    \textnormal{Tr} \,\Lambda^l \Lambda^m \;=\; (C_l)^2 \, l (l+1) \,
    \delta^{lm} \;=\; 2 \,\delta^{lm} \,.
\end{equation}

\emph{Proof of condition \eqref{proofprop8}.} Analogously to the proofs \eqref{proofss}
and \eqref{proofaa} we find
\begin{equation}
    \textnormal{Tr} \,\Lambda^{jk}_a \Lambda^{mn}_s \;=\; i \left(
    - \,\delta^{jn} \delta^{km} \,+\, \delta^{jm} \delta^{kn} \,-\, \delta^{jm}
    \delta^{kn} \,+\, \delta^{jn} \delta^{km} \right) \;=\; 0 \,.
\end{equation}

\emph{Proof of condition \eqref{proofprop9}.} Inserting definitions \eqref{ggms} and
\eqref{ggmd} gives
\begin{eqnarray}
    \text{Tr} \,\Lambda^{jk}_s \Lambda^m & \;=\; & C_m \sum_{p=1}^d \Big(
        - m \langle p | k \rangle \langle j | m+1 \rangle \langle m+1 | p
        \rangle \,-\, m \langle p | j \rangle \langle k | m+1 \rangle
        \langle m+1 | p \rangle \nonumber\\
    & & + \sum_{n=1}^{m} \langle p | j \rangle \langle k | n \rangle
        \langle n | p \rangle \,+\, \sum_{n=1}^m \langle p | k \rangle
        \langle j | n \rangle \langle n | p \rangle \Big)
        \nonumber\\
    & \;=\; & - \,2m \,\delta^{j (m+1)} \delta^{k (m+1)} \,+\, 2 \sum_{l=1}^m
    \delta^{kl} \delta^{jl} \nonumber\\
    & \;=\; & 0 \,,
\end{eqnarray}
since per definition we have $j < k\,$.\\

\emph{Proof of condition \eqref{proofprop10}.} This proof is equivalent to the previous
one since constant factors in front of the terms do not matter.

\subsection{Calculation of term B in GGB}\label{termB}

To obtain the Bloch vector notation of term $B$ \eqref{isoggmB} we insert the standard
matrix expansion \eqref{smggb} for the case $j = k$. We split the tensor products in the
following way
\begin{equation}
    B \;=\; \frac{1}{d} \left( B_1 \,+\, B_2 \,+\, B_3 \,+\, B_4 \,+\,
    \frac{1}{d} \,\mathbbm{1} \otimes \mathbbm{1} \right) \,,
\end{equation}
where the terms $B_1, \ldots ,B_4$ are introduced by (note that $\Lambda^0 = 0$)
\begin{eqnarray}
    B_1 & \;=\; & \sum_{j=1}^d \left( \frac{j-1}{2j} \Lambda^{j-1} \otimes
        \Lambda^{j-1} \,+\, \sum_{n(=l)=0}^{d-j-1} \frac{1}{2(j+n)(j+n+1)} \Lambda^{j+n}
        \otimes \Lambda^{j+n} \right) \label{isoggbb1} \\
    B_2 & \;=\; & \sum_{j=1}^d \Bigg( - \sum_{l=0}^{d-j-1} \sqrt{\frac{j-1}
        {4j(j+l)(j+l+1)}} \Lambda^{j-1} \otimes \Lambda^{j+l}
        \nonumber\\
        & & \qquad \ \, - \sum_{n=0}^{d-j-1} \sqrt{\frac{j-1}
        {4j(j+n)(j+n+1)}} \Lambda^{j+n} \otimes \Lambda^{j-1} \nonumber\\
        & & \qquad \ \, + \sum_{n \neq l,\, n,l=0}^{d-j-1} \frac{1}{2
        \sqrt{(j+n)(j+n+1)(j +
        l)(j+l+1)}} \Lambda^{j+n} \otimes \Lambda^{j+l} \Bigg)
        \label{isoggbb2} \\
    B_3 & \;=\; & \frac{1}{d} \sum_{j=1}^d \left( - \sqrt{\frac{j-1}{2j}}
        \Lambda^{j-1} \otimes \mathbbm{1} \,+\, \sum_{n=0}^{d-j-1}
        \frac{1}{\sqrt{2(j+n)(j+n+1)}} \Lambda^{j+n} \otimes
        \mathbbm{1} \right) \label{isoggb3} \\
    B_4 & \;=\; & \frac{1}{d} \sum_{j=1}^d \left( - \sqrt{\frac{j-1}{2j}}
        \mathbbm{1} \otimes \Lambda^{j-1} \,+\, \sum_{l=0}^{d-j-1}
        \frac{1}{\sqrt{2(j+l)(j+l+1)}} \mathbbm{1} \otimes
        \Lambda^{j+l} \right) \label{isoggb4} \,.
\end{eqnarray}
Only the first term $B_1$ \eqref{isoggbb1} gives a contribution
\begin{equation}
    B_1 \;=\; \sum_{m=1}^{d-1} \left( \frac{m}{2(m+1)} \,+\,
    \frac{m}{2m(m+1)} \right) \Lambda^m \otimes \Lambda^m \;=\; \frac{1}{2}
    \sum_{m=1}^{d-1} \Lambda^m \otimes \Lambda^m \,,
\end{equation}
whereas the remaining terms vanish:
\begin{eqnarray}
    B_2 & \;=\; & \sum_{m < p,\, m,p =1}^{d-1} \left( -
        \sqrt{\frac{m}{4(m+1)p(p+1)}} \,+\, \frac{m}{\sqrt{4m(m+1)p(p+1)}}
        \right) \Lambda^m \otimes \Lambda^p \nonumber\\
    & & + \sum_{m > p,\, m,p =1}^{d-1} \left( -
        \sqrt{\frac{p}{4(p+1)m(m+1)}} \,+\, \frac{p}{\sqrt{4p(p+1)m(m+1)}}
        \right) \Lambda^m \otimes \Lambda^p \nonumber\\
    & \;=\; & \left( \sum_{m<p} \frac{-m+m}{2 \sqrt{m(m+1)p(p+1)}} \,+\,
        \sum_{m>p} \frac{-p+p}{2 \sqrt{m(m+1)p(p+1)}} \right) \Lambda^m
        \otimes \Lambda^p \nonumber\\
    & \;=\; & 0 \,,
\end{eqnarray}
and in quite the same manner
\begin{eqnarray}
    B_3 & \;=\; & \frac{1}{d} \sum_{m=1}^{d-1}
        \frac{-m+m}{\sqrt{2m(m+1)}} \ \Lambda^m \otimes \mathbbm{1} = 0 \,, \nonumber\\
    B_4 & \;=\; & \frac{1}{d} \sum_{p=1}^{d-1}
        \frac{-p+p}{\sqrt{2p(p+1)}} \ \mathbbm{1} \otimes \Lambda^p = 0 \,.
\end{eqnarray}
Thus we find the following Bloch vector of $B$ \eqref{isoggmB}
\begin{equation}
    B \;=\; \frac{1}{2d}
    \sum_{m=1}^{d-1} \Lambda^m \otimes \Lambda^m \,+\, \frac{1}{d^2} \,\mathbbm{1} \otimes
    \mathbbm{1} \,.
\end{equation}

\subsection{Proof of Orthonormality of WOB} \label{secproofon}

For proofs relevant in the WOB we often need the equivalence
\begin{equation} \label{complexsumrule}
    \sum_{n=0}^{d-1} e^{\frac{2 \pi i}{d} \,nx} \;=\; \begin{cases}
        d & \text{if } x=0 \\
        0 & \text{if } x\neq 0
    \end{cases}, \quad x \in \mathbbm{Z} \,.
\end{equation}
So we use Eq.~\eqref{complexsumrule} to proof the orthonormality \eqref{orthonormwo} of
the Weyl operators \eqref{defwo}
\begin{eqnarray}
    \text{Tr} \,U_{nm}^{\dag} U_{lj} & \;=\; & \sum_{p=0}^{d-1} \sum_{k,
        \tilde{k} = 0}^{d-1} e^{\frac{2 \pi i}{d}\,(\tilde{k}l-kn)} \,\langle p |
        (k+m) \,\textrm{mod}\,d \rangle \langle k | \tilde{k} \rangle \langle
        (\tilde{k} + j ) \,\textrm{mod}\,d |p \rangle \nonumber\\
    & \;=\; & \sum_{p=0}^{d-1} \sum_{k,
        \tilde{k} = 0}^{d-1} e^{\frac{2 \pi i}{d}\,(\tilde{k}l-kn)} \,\langle p |
        (k+m) \,\textrm{mod}\,d \rangle  \langle
        (\tilde{k} + j ) \,\textrm{mod}\,d |p \rangle \,\delta_{k \tilde{k}} \nonumber\\
    & \;=\; & \sum_{k=0}^{d-1} e^{\frac{2 \pi i}{d}\,k (l-n)} \,\delta_{mj}
        \nonumber\\
    & \;=\; & d \,\delta_{nl} \,\delta_{mj} \,.
\end{eqnarray}

\subsection{Expansion into WOB}\label{BellWOB}

Formula \eqref{maxentwob2} for the Bell state in terms of WOB we derive in the following
way. We express the standard matrices by the WOB \eqref{smwob}, rewrite the indices and
separate the nonvanishing terms
\begin{eqnarray} \label{maxentwob1}
  \left| \phi_+^d \right\rangle \left\langle \phi_+^d \right|
    & \;=\; & \frac{1}{d} \sum_{j,k =1}^{d} |j \rangle \langle k |
        \otimes |j \rangle \langle k| \nonumber\\
    & \;=\; & \frac{1}{d^3} \sum_{j,k=0}^{d-1} \sum_{l,l'=0}^{d-1} e^{-\frac{2 \pi i}{d}\,j
        (l+l')} U_{l (k-j) mod \, d} \otimes U_{l' (k-j) mod \, d}
        \nonumber\\
    & \;=\; & \frac{1}{d^3} \sum_{m,k=0}^{d-1} \sum_{l,l'=0}^{d-1} e^{-\frac{2 \pi i}{d}\,
        (k-m)(l+l')} U_{l m} \otimes U_{l' m} \nonumber\\
    & \;=\; & \frac{1}{d^2} \left( \sum_m U_{0m} \otimes U_{0m} \,+\,
        \sum_m \, \sum_{l,l';\,l+l'=d} U_{lm} \otimes U_{l'm} \right) \nonumber\\
    & & +\; \frac{1}{d^3} \sum_m \, \sum_{
        l,l';\,l,l' \neq 0,0;\,l+l' \neq d} \left( \sum_k e^{-\frac{2 \pi i}{d}\,(k-m) (l+l')}
       \right) U_{lm} \otimes U_{l'm} \,.
\end{eqnarray}
The last term in Eq.~\eqref{maxentwob1} vanishes due to relation \eqref{complexsumrule}.
Identifying $U_{00} = \mathbbm{1}$ and using the notation with negative values of the
index $l$, which have to be considered as $mod \ d\,$, we gain the formula
\begin{equation}
    \left| \phi_+^d \right\rangle \left\langle \phi_+^d \right| \;=\;
    \frac{1}{d^2} \,\mathbbm{1} \otimes \mathbbm{1} \,+\, \frac{1}{d^2}
    \,\sum_{l,m = 0}^{d-1} U_{lm} \otimes U_{-lm} \,, \qquad (l,m) \neq (0,0) \,.
\end{equation}

\bibliography{references}

\end{document}